\documentclass[twocolumn]{aastex631}
\usepackage{mathtools}

\usepackage{orcidlink}
\usepackage{xcolor}
\usepackage{array}
\usepackage{multirow}
\usepackage{tablefootnote}
\renewcommand{\arraystretch}{16}

\shorttitle{The Inconsistent use of $\omega$ in the RV Equation}
\shortauthors{Householder \& Weiss}

\begin{document}

\title{The Inconsistent use of $\omega$ in the RV Equation}

\author{Aaron Householder \orcidlink{0000-0002-5812-3236}}
\affiliation{Department of Astronomy, Yale University,
52 Hillhouse, New Haven, CT 06511, USA}
\author{Lauren Weiss \orcidlink{0000-0002-3725-3058}}
\affiliation{Department of Physics and Astronomy, University of Notre Dame, Notre Dame, IN 46556, USA}

\begin{abstract}
Since the discovery of the first exoplanet orbiting a main-sequence star, astronomers have inferred the orbital properties of planets using stellar radial velocity (RV) measurements. For a star orbited by a single planet, the stellar orbit is a dilation and $180^\circ$ rotation of the planetary orbit. Thus, many of the Keplerian orbital properties of the star are identical to those of the planet. However, there is a notable exception: the argument of periastron, $\omega$, defined as the angle between the periapsis of an orbiting body and its ascending node. The argument of periastron of the star ($\omega_\star$) is $180^\circ$ offset from the argument of periastron of the planet ($\omega_p$). This distinction is important because some derivations of the RV equation use $\omega_\star$, while others use $\omega_p$. This discrepancy arises because commonly used derivations of the RV equation do not adhere to a single coordinate system. As a result, there are inconsistencies in the definitions of the Keplerian orbital parameters in various RV models, leading to values of the ascending node and $\omega$ that are $180^\circ$ offset. For instance, some packages, such as \texttt{RadVel} and \texttt{ExoFast}, report values for $\omega_{\star}$ that are identical to the $\omega_p$ values determined with other packages, such as \texttt{TTVFast} and \texttt{Orvara}, resulting in orbital solutions that differ by $180^\circ$. This discrepancy highlights the need for standardized conventions and definitions in RV modeling, particularly as we enter the era of combining RVs with astrometry.

\end{abstract}

\section{Introduction}
\label{Section1}

The radial velocity (RV) method is one of the most common ways to detect and characterize exoplanets. According to the NASA Exoplanet Archive (accessed 2022 Nov 07), the RV method is responsible for the characterization of nearly 2000 exoplanets \citep{2013PASP..125..989A}. The RV method works by measuring the Doppler shift in the spectral features of a star, which allows us to determine changes in the stellar line of sight velocity caused by the gravitational influence of orbiting planets. This technique gives us information about planetary masses as well as the orbital properties of exoplanets and their host stars.

When modeling RVs, we must model the argument of periastron, $\omega$, which is measured from the ascending node to the periastron, in the direction of a body's orbit. In other words, $\omega$ describes the orientation of a body's elliptical path within the orbital plane. For a star-planet system, the argument of periastron for the star ($\omega_\star$) is offset by $180^\circ$ from the argument of periastron of the planet ($\omega_p$). However, there are two inconsistent interpretations of the RV equation, one with $\omega_p$ and the other with $\omega_\star$. In this paper, we demonstrate that this discrepency is a consequence of the use of inconsistent coordinate systems to model RVs. We also develop a test from first-principles that highlights this discrepancy and demonstrate that various packages that model RVs report $\omega$ inconsistently.

\section{The RV Equation: $\omega_p$ versus $\omega_\star$}
\label{Section2}
Although the RV method is commonly used for exoplanet detection and characterization, there is disagreement across various sources regarding which equation to use, with some using $\omega_p$ and others using $\omega_\star$. To address this discrepancy, we derive the RV equation, following the approach outlined in many modern sources \citep{1999ssd..book.....M,2010exop.book...27L}. First, we consider a system with a star of mass $M_{\star}$ and a planet of mass $m_p$. For RV measurements, we observe the motion of the star around the barycenter. The position of the star relative to the barycenter can be described using three orbital elements: the semi-major axis of the star (a), the eccentricity (e), and the true anomaly ($\nu$):

\begin{equation}
\label{eq1}
r_{\star}=\left(\frac{m_p}{M_{\star}+m_p}\right) \frac{a\left(1-e^2\right)}{1+e \cos \nu}
\end{equation}

It is useful to express the position of the star in a Cartesian coordinate system ($x_{\star}$, $y_{\star}$, and $z_{\star}$) centered at the barycenter. The x-axis points towards the pericenter of the orbit, the y-axis lies in the orbital plane, and the z-axis is mutually perpendicular to both the x and y axes, forming a right-handed triad. The position of the star is given by the following: 

\begin{equation}
\left(\begin{array}{c}
x_{\star} \\ 
y_{\star} \\ 
z_{\star} 
\end{array}\right) =
\left(\begin{array}{c}
r_{\star} \cos \nu \\
r_{\star} \sin \nu\\
0 
\end{array}\right)
\end{equation}

We can also define a sky-plane coordinate system ($X_{\star}$, $Y_{\star}$, $Z_{\star}$) with the barycenter as the origin and $X_{\star}$ pointing North, $Y_{\star}$ pointing East (in the direction of increasing right ascension), and $Z_{\star}$ pointing toward the observer. To transform the coordinates $x_{\star}$, $y_{\star}$, and $z_{\star}$ into the sky-plane, we apply three rotations about the orbital plane through the angles $\omega_{\star}$, $I$, and $\Omega$. To avoid confusion, we explicitly define the meanings of these angles and illustrate their relationships in Figure \ref{Fig1}. In this paper, the ascending node refers to the point where the stellar orbit intersects the reference plane, moving from ``below''to ``above'', or when the star crosses the line of nodes from $Z < 0$ to $Z > 0$ \citep{1999ssd..book.....M,2010exop.book...15M}. This definition of the ascending node is not universally accepted, leading to confusion that will be further discussed in later portions of this manuscript. The longitude of the ascending node, $\Omega$, is the angle between the reference line ($X_\star$) and the ascending node. The argument of periastron of the star, $\omega_{\star}$, is the angle between the ascending node and the stellar periapsis, in the direction of the stellar orbit. The inclination, $I$, is the angle orbital plane and the reference plane. Using these definitions, we can transform $x_{\star}$, $y_{\star}$, and $z_{\star}$ into the sky plane with the three coordinate transformations given in \citet{1999ssd..book.....M}. Thus, the position of the star in the sky plane ($X_{\star}$, $Y_{\star}$, $Z_{\star}$) is expressed as a function of the true anomaly $\nu$:

\begin{figure}
    \centering
    \includegraphics[width=0.5\textwidth]{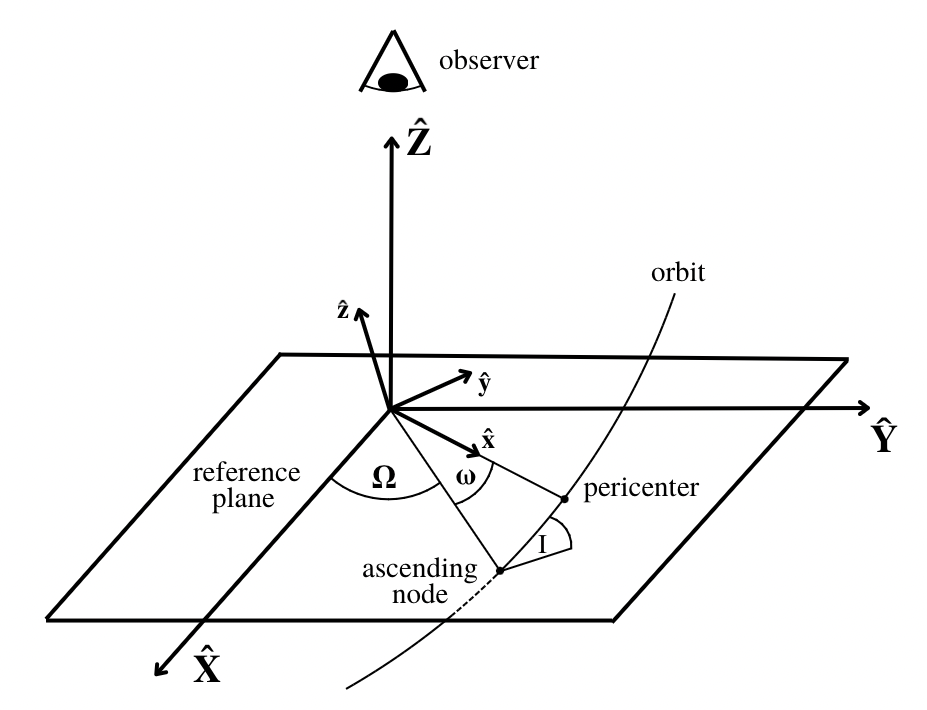}
    \caption{The 3-dimensional motion of an orbit with respect to the reference plane, re-produced from Figure 4 of \citet{2010exop.book...15M}.}
    \label{Fig1}
\end{figure}

\begin{small}
\begin{equation}
\label{eq3}
\left(\begin{array}{c}
r_{\star}\left[\cos \Omega \cos \left(\omega_{\star}+\nu\right)-\sin \Omega \sin \left(\omega_{\star}+\nu\right) \cos I\right] \\
r_{\star}\left[\sin \Omega \cos \left(\omega_{\star}+\nu\right)+\cos \Omega \sin \left(\omega_{\star}+\nu\right) \cos I\right] \\
r_{\star}\left[\sin \left(\omega_{\star}+\nu\right) \sin I\right]
\end{array}\right)
\end{equation}
\end{small}

 To calculate the radial velocity of the star, RV, we can differentiate $Z_{\star}$ with respect to time: 
 
\begin{equation}
\label{eq4}
RV = \dot{Z}_{\star} = K[\cos (\nu+\omega_\star)+e \cos (\omega_\star)] 
\end{equation}

\noindent where $K$ is the RV semi-amplitude. 

This is the same equation derived in \citet{2010exop.book...27L}. Similarly to the derivation above, \citet{2010exop.book...27L} derive the RV equation with $\hat{Z}$ pointing towards the observer (confirmed via personal communication, \textit{Lovis 2022}). A common source of confusion is that the \citet{2010exop.book...27L} RV equation (Equation \ref{eq4}) does not follow the standard observational convention for RV measurements. A positive RV corresponds to a stellar blue-shift in the \citet{2010exop.book...27L} RV equation. However, observational astronomers typically choose the convention where RV is positive for stellar red-shifts \citep{2018PASP..130d4504F, 2014ApJ...787..132D, 2013PASP..125...83E}. Therefore, if we want to follow the standard observational convention, we cannot use the coordinate system and RV equation presented in \citet{2010exop.book...27L}\footnote{It is worth noting that \citet{2010exop.book...27L} switch $\hat{Z}$ to point away from the observer in subsequent sections of the text: opposite the coordinate system used in their derivation of the RV equation (confirmed via personal communication, \textit{Lovis 2022}).}.

As a result, there are inconsistent ways of modeling RVs throughout the literature. One approach, used in packages like \texttt{TTVFast} \citep{2014ApJ...787..132D} and \texttt{Orvara} \citep{2021AJ....162..186B}, is to adopt the same coordinate system as \citet{2010exop.book...27L}, where $\hat{Z}$ points towards the observer. In this coordinate system, we must introduce a negative sign to the \citet{2010exop.book...27L} RV equation to follow the standard observational convention: 
 
\begin{equation}
\label{Eq5}
RV =-\dot{Z}_{\star} =-K[\cos (\nu+\omega_\star)+e \cos (\omega_\star)]
\end{equation}
Equivalently, Equation \ref{Eq5} can be re-written in terms of $\omega_p$ rather than $\omega_\star$ (given that $\omega_\star$ = $\omega_p$ + $180^\circ$): 
\begin{equation}
\label{Eq6}
RV= -\dot{Z}_{\star} =K[\cos (\nu+\omega_p)+e \cos (\omega_p)]
\end{equation}

Other sources, such as \texttt{RadVel} \citep{2018PASP..130d4504F} and \texttt{ExoFast} \citep{2013PASP..125...83E}, take a different approach by using a coordinate system where $\hat{Z}$ points away from the observer. In this coordinate system, a positive RV corresponds to a stellar red-shift, so the RV equation becomes:

\begin{equation}
\label{Eq9}
RV =\dot{Z}_{\star} = K[\cos (\nu+\omega_\star)+e \cos (\omega_\star)] 
\end{equation}

The reason that \texttt{RadVel} and \texttt{ExoFast} do not use $\omega_p$ is that switching $\hat{Z}$ to point away from the observer also produces a different position of the ascending node. If we define the ascending node as crossing from $Z<0$ to $Z>0$, then the position of the ascending node is dependent on the direction of $\hat{Z}$:

\begin{figure*}
    \centering
    \includegraphics[width=0.95\textwidth]{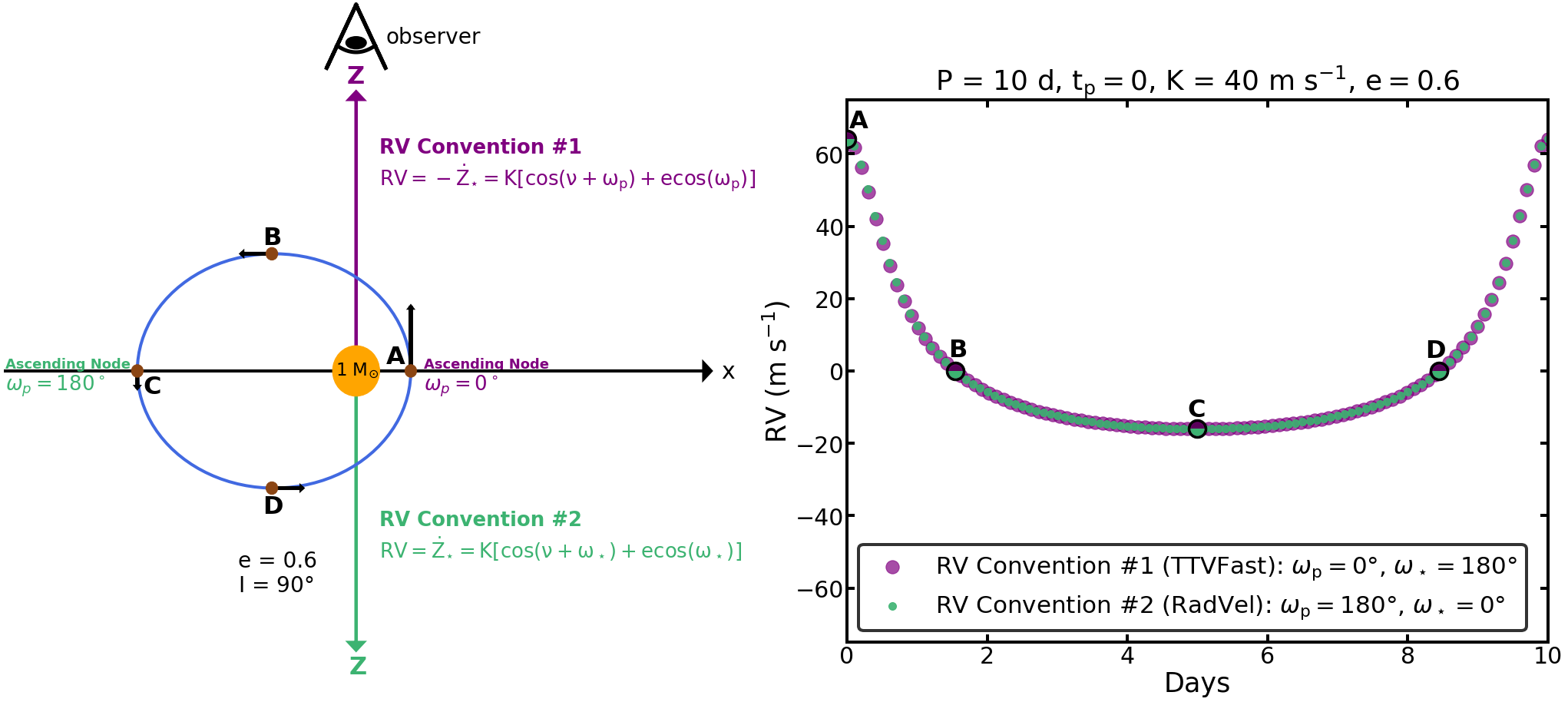}
    \caption{Left: A 2-dimensional depiction of an exoplanet (brown) orbiting a star (orange). The planetary system has the following properties: $M_* = 1$ M$_{\odot}$, $P = 10$ days, $K = 40 $ m s$^{-1}$, $M_\mathrm{p} = 107.8$ M$_{\oplus}$, $e = 0.6$, $I = 90^\circ$, $t_\mathrm{p} = 0 $. Labels A, B, C, and D denote four successive phases of the planetary orbit. For instance, at A, the planet is at periastron moving directly towards the observer, so the star is moving directly away from the observer (i.e. a positive RV). Notably, the different RV conventions yield different values of $\omega_p$ for the same orbital motion. This is because the position of the ascending node in convention 1 ($\hat{Z}$ toward observer, purple) is $180^\circ$ offset from the ascending node in convention 2 ($\hat{Z}$ away from observer, green). Right: The radial velocity of the star produced from the configuration and orientation of the planetary system on the left generated with $\texttt{TTVFast}$ (convention 1, purple) and $\texttt{RadVel}$ (convention 2, green). If $\omega$ were modeled consistently, the $180^\circ$ difference in $\omega$ values should cause the RV curve produced by $\texttt{RadVel}$ to be inverted compared to the RV curve generated by $\texttt{TTVFast}$. Thus, the fact that \texttt{Radvel} and \texttt{TTVFast} yield the same RV curve despite different inputs of $\omega$ demonstrates the inconsistent modeling of $\omega$ with different RV conventions. 
    } \label{Fig2}
\end{figure*}

\begin{itemize}
\item $\hat{Z}$ towards the observer: The ascending node is the point when the orbit of a body intersects the line of nodes and the body is approaching the observer (blue-shifted)  (i.e. \citealt{1999ssd..book.....M,2010exop.book...15M,2010exop.book...27L}). 

\item $\hat{Z}$ away from the observer: The ascending node is the point when the orbit of a body intersects the line of nodes and the body is moving away from the observer (red-shifted) (i.e. \citealt{1913PASP...25..208P, 2018PASP..130d4504F, 2013PASP..125...83E}). 
\end{itemize}

Therefore, the ascending node (and by extension the value of $\omega$) in the \texttt{RadVel} and \texttt{ExoFast} coordinate system is $180^\circ$ offset from the ascending node in the \texttt{Orvara} and \texttt{TTVFast} coordinate system. Thus, \texttt{RadVel} and \texttt{ExoFast} effectively introduce two negative signs to the \citet{2010exop.book...27L} RV equation ($\hat{Z}$ away from observer; ascending node $180^\circ$ offset), which yields an RV equation identical to the \citet{2010exop.book...27L} RV equation (with $\omega_\star$)\footnote{Some planetary dynamics texts, such as  \citet{2023dyps.book.....T}, derive the RV of the star in terms of the orbital properties of the planet: $RV = -\dot{Z}_{p} = -K[\cos (\nu+ \omega_p)+e \cos (\omega_p)]$. This is the same convention used in \texttt{RadVel} and \texttt{ExoFast} with the following substitutions: $\dot{Z}_{\star} = -\dot{Z}_{p}$, $\omega_\star$ = $\omega_p$ + $180^\circ$.}, albeit with much different conventions.

It is worth noting that there are additional RV conventions in the literature that are not thoroughly discussed here. This is because the ascending node is not exclusively interpreted as crossing from $Z < 0$ to $Z > 0$.  For example, \texttt{Exoplanet} \citep{2013PASP..125..306F} uses a coordinate system with $\hat{Z}$ pointed towards the observer but defines the ascending node as the point at which the orbit intersects the sky plane while moving away from the observer (red-shifted). As a result, a body at the ascending node in \texttt{Exoplanet} crosses from $Z > 0$ to $Z < 0$, unlike our previously defined convention of crossing from  $Z < 0$ to $Z > 0$. If we define the ascending node in terms of Doppler-shifts like \texttt{Exoplanet}, the main difference is that the position of the ascending node is independent of $\hat{Z}$: Doppler shifts do not change based on the direction of the Z-axis. However, not all sources use a red-shifted definition of the ascending node. For instance, in many textbooks, (i.e. \citealt{2010exop.book...15M, 1999ssd..book.....M}), a body intersects the sky plane while approaching the observer (a blue-shift) at the ascending node. Thus, even if we define the ascending node in terms of Doppler shifts, the inconsistency in the RV equation still persists. However, the reason for the discrepancy would stem from the difference in red-shifted (i.e. \citealt{1913PASP...25..208P,2013PASP..125..306F,2013PASP..125...83E,2018PASP..130d4504F}) versus blue-shifted (i.e. \citealt{1999ssd..book.....M,2010exop.book...15M,2010exop.book...27L,2014ApJ...787..132D,2021AJ....162..186B}) definitions of the ascending node, which is equivalent to a $180^\circ$ difference in $\omega$.

\section{Discrepancy in modeling $\omega$}
We developed a test  to determine the effect inconsistent coordinate systems has for modeling RVs (Figure \ref{Fig1}). First, we built a mock star-planet system for a 107.8 M$_{\oplus}$ planet orbiting a 1 M$_{\odot}$ star using \texttt{TTVFast}. The orbiting planet has the following properties: $P = 10$ days, $e = 0.6$, $I= 90^\circ$, $\omega_p = 0 ^\circ$ (i.e. $\omega_\star = 180^\circ$), the longitude of the ascending node $\Omega = 0^\circ$, and the mean anomaly $M = 0^\circ$ (i.e. the time of periastron passage $t_\mathrm{p} = 0 $). These parameters were chosen to induce a RV semi-amplitude of $ K = 40$\,m $\mathrm{s^{-1}}$. The planet in our mock system has an argument of periastron of $ 0^\circ$. This implies that the planet is at periastron as it crosses the reference plane and passes through the ascending node. Therefore, since $\hat{Z}$ points towards the observer in \texttt{TTVFast}, at t = 0, we expect the planet to be at pericenter moving towards the observer at maximum absolute velocity. At that time, the star should also be at pericenter moving away from the observer at maximum absolute velocity. $\texttt{TTVFast}$ follows the standard observational convention where RV is positive when the star moves away from the observer. Therefore, at t = 0, the star should be at maximum radial velocity. This expected behavior agrees with our synthetic RVs from \texttt{TTVFast} (see Figure \ref{Fig2}). 

Next, we used \texttt{RadVel} to generate synthetic RVs for an input of the following parameters: $P = 10$ days, $t_\mathrm{p} = 0$, $e = 0.6$, $\omega_p = 180^\circ$ (i.e. $\omega_{\star} = 0^\circ$), and $ K = 40$\,m $\mathrm{s^{-1}}$. These are the same input parameters we used for \texttt{TTVFast} but we changed $\omega_p$ by $180^\circ$ ($\Omega$ in \texttt{TTVFast} is a rotational angle on the projected sky plane, so it does not affect the expected RV curve). Despite the $180^\circ$ difference in $\omega$, \texttt{RadVel} computes the same stellar RV curve as \texttt{TTVFast} (see Figure \ref{Fig2}).  Thus, the value of $\omega$ alone (even when specified as $\omega_\star$ or $\omega_p$) is insufficient to uniquely determine the RV of the star.  The RV convention (including the direction of $\hat{Z}$ and the location of the ascending node) must also be specified for complete clarity. 

\section{Scope}
It is difficult to determine the scope of this issue because many papers that model RVs do not specify whether they report $\omega_{\star}$ or $\omega_p$ 
but report the ambiguous $\omega$ instead. Nonetheless, the inconsistent modeling of $\omega$ has resulted in varied reporting of $\omega$ values throughout the literature. For example, we fit the RVs of Pi Men from \citet{2006ApJ...646..505B} with four different RV modeling packages (\texttt{ExoFast}, \texttt{Orvara}, \texttt{RadVel}, and \texttt{TTVFast}), and the resulting values of $\omega_\star$ varied by $180^\circ$ (see Table \ref{table:1}). 

It will be crucial to address the discrepancy in $\omega$ modeling as we prepare to combine archival RVs with astrometry in GAIA DR4. This inconsistency could also significantly impact our ability to identify the most suitable targets for direct-imaging observations, many of which will likely be RV-detected planets. For instance, for Pi Men b, \citet{2021A&A...651A...7C} finds that the value of $\omega_\star$ used strongly effects the planet-star-observer phase angle, $\alpha_{o b s}$. For an $\omega_{\star}$ value similar to \texttt{RadVel} and \texttt{ExoFast}, \citet{2021A&A...651A...7C} finds that $\alpha_{o b s} = \left[42_{-3}^{+16}, 111_{-7}^{+2}\right]$, whereas for an $\omega_{\star}$ value similar to \texttt{TTVFast} and \texttt{Orvara}, the range of $\alpha_{o b s}$ decreases to $\left[69_{-2}^{+7}, 95_{-1}^{+1}\right]$. Thus, inconsistencies in $\omega_{\star}$ could potentially lead to incorrect selection or exclusion of Pi Men b for phase-curve measurements.

\begin{table*}
\caption{The coordinate system, RV equation, and the argument of periastron of the star ($\omega_\star$) for Pi Men b using five different sources: \citet{2010exop.book...27L},  \texttt{Orvara}, \texttt{TTVFast}, \texttt{RadVel}, and \texttt{ExoFast}. \texttt{Orvara} (confirmed via personal communication, \textit{Brandt 2023}) and \texttt{TTVFast} both model $\omega_p$, but their $\omega_\star$ values are reported here for a clear comparison. It is also worth noting that the RV equation and coordinate system in \citet{2010exop.book...27L} defines RVs as positive for stellar blue-shifts (confirmed via personal communication, \textit{Lovis 2022}). This is in contrast to the convention used by most RV astronomers, where a positive RV corresponds to a stellar red-shift.
}

\centering
\begin{tabular}{c|c|c|c}
\hline
\textbf{Source} & $\hat{\textbf{Z}}$ & \textbf{RV Equation} & $\omega_\star$ \\
\hline
\hline
\citet{2010exop.book...27L} & Toward the Observer & $RV = \phantom{-}\dot{Z} = K[\cos (\nu+\omega_\star)+e \cos (\omega_\star)]$& - \\
\hline
\texttt{TTVFast}  & Toward the Observer & $RV = -\dot{Z} = K[\cos (\nu+\omega_p)+e \cos (\omega_p)]$ & $151.2 \pm 1.4 ^\circ$ \\
\hline
\texttt{Orvara} & Toward the Observer & $RV = -\dot{Z} = K[\cos (\nu+\omega_p)+e \cos (\omega_p)]$ &$151.3 \pm 1.0 ^\circ$ \\
\hline
\texttt{ExoFast}  & Away From Observer & $RV = \phantom{-}\dot{Z} = K[\cos (\nu+\omega_\star)+e \cos (\omega_\star)]$& $331.1 \pm 1.0^\circ $\\
\hline
\texttt{RadVel}  & Away From Observer & $RV = \phantom{-}\dot{Z} = K[\cos (\nu+\omega_\star)+e \cos (\omega_\star)]$ &  $331.2 \pm 1.2^\circ$\\
\hline
\end{tabular}
\label{table:1}
\end{table*}

\section{Conclusion}
Here, we provide a summary of the findings of this manuscript in bullet form: 

\begin{itemize}
\item \citet{2010exop.book...27L} derive an RV equation with $\hat{Z}$ pointing towards observer where a positive RV corresponds to a stellar blue-shift. 

\item To follow the standard observational convention where a positive RV corresponds to a stellar redshift, many sources, such as \texttt{TTVFast} and \texttt{Orvara}, introduce a negative sign to the \citet{2010exop.book...27L} RV equation  (effectively replacing $\omega_{\star}$ with $\omega_p$).

\item Other sources, such as \texttt{RadVel} and \texttt{ExoFast}, use a different direction of $\hat{Z} $ than \citet{2010exop.book...27L}, leading to an ascending node that is $180^\circ$ offset. This results in an RV equation with $\omega_{\star}$ rather than $\omega_p$. 

\item Therefore, there are two frequently used RV equations and coordinate systems that are inconsistent with each other: one with $\omega_p$ and $\hat{Z}$ pointing towards the observer, and the other with $\omega_{\star}$ and $\hat{Z}$ pointing away from the observer.

\end{itemize}

In future work, we encourage authors to be explicit in defining their coordinate system, the ascending node, and their use of $\omega_p$ or $\omega_\star$. This is especially relevant as we approach Gaia DR4, which will allow us to constrain $\omega$ better than ever before. 

AH acknowledges support from the National Science Foundation REU Program (grant no. 2050527). LMW acknowledges support from the NASA Exoplanet Research Program through grant 80NSSC23K0269.  Additionally, AH thanks Andrew Mayo and Greg Laughlin for valuable discussions, as well as the members of the AstroWeiss group for their helpful feedback on plotting. AH also thanks the anonymous referee and Christoph Lovis for their valuable and insightful comments. Furthermore, AH thanks Jason Eastman for serving as a referee, and the \texttt{RadVel} team for their understanding and feedback during this process.

\software{\texttt{ExoFast} \citep{2013PASP..125...83E},
\texttt{Orvara} \citep{2021AJ....162..186B},
\texttt{RadVel} \citep{2018PASP..130d4504F}, \texttt{TTVFast} \citep{2014ApJ...787..132D},  \texttt{NumPy} \citep{numpy}, \texttt{Matplotlib} \citep{matplotlib}, \texttt{Pandas} \citep{pandas}
}
\bibliography{references}{}

\begin{thebibliography}{}
\expandafter\ifx\csname natexlab\endcsname\relax\def\natexlab#1{#1}\fi
\providecommand{\url}[1]{\href{#1}{#1}}
\providecommand{\dodoi}[1]{doi:~\href{http://doi.org/#1}{\nolinkurl{#1}}}
\providecommand{\doeprint}[1]{\href{http://ascl.net/#1}{\nolinkurl{http://ascl.net/#1}}}
\providecommand{\doarXiv}[1]{\href{https://arxiv.org/abs/#1}{\nolinkurl{https://arxiv.org/abs/#1}}}

\bibitem[{{Akeson} {et~al.}(2013){Akeson}, {Chen}, {Ciardi}, {Crane}, {Good},
  {Harbut}, {Jackson}, {Kane}, {Laity}, {Leifer}, {Lynn}, {McElroy}, {Papin},
  {Plavchan}, {Ram{\'\i}rez}, {Rey}, {von Braun}, {Wittman}, {Abajian}, {Ali},
  {Beichman}, {Beekley}, {Berriman}, {Berukoff}, {Bryden}, {Chan}, {Groom},
  {Lau}, {Payne}, {Regelson}, {Saucedo}, {Schmitz}, {Stauffer}, {Wyatt}, \&
  {Zhang}}]{2013PASP..125..989A}
{Akeson}, R.~L., {Chen}, X., {Ciardi}, D., {et~al.} 2013, \pasp, 125, 989,
  \dodoi{10.1086/672273}

\bibitem[{{Brandt} {et~al.}(2021){Brandt}, {Dupuy}, {Li}, {Brandt}, {Zeng},
  {Michalik}, {Bardalez Gagliuffi}, \& {Raposo-Pulido}}]{2021AJ....162..186B}
{Brandt}, T.~D., {Dupuy}, T.~J., {Li}, Y., {et~al.} 2021, \aj, 162, 186,
  \dodoi{10.3847/1538-3881/ac042e}

\bibitem[{{Butler} {et~al.}(2006){Butler}, {Wright}, {Marcy}, {Fischer},
  {Vogt}, {Tinney}, {Jones}, {Carter}, {Johnson}, {McCarthy}, \&
  {Penny}}]{2006ApJ...646..505B}
{Butler}, R.~P., {Wright}, J.~T., {Marcy}, G.~W., {et~al.} 2006, \apj, 646,
  505, \dodoi{10.1086/504701}

\bibitem[{{Carri{\'o}n-Gonz{\'a}lez} {et~al.}(2021){Carri{\'o}n-Gonz{\'a}lez},
  {Garc{\'\i}a Mu{\~n}oz}, {Santos}, {Cabrera}, {Csizmadia}, \&
  {Rauer}}]{2021A&A...651A...7C}
{Carri{\'o}n-Gonz{\'a}lez}, {\'O}., {Garc{\'\i}a Mu{\~n}oz}, A., {Santos},
  N.~C., {et~al.} 2021, \aap, 651, A7, \dodoi{10.1051/0004-6361/202039993}

\bibitem[{{Deck} {et~al.}(2014){Deck}, {Agol}, {Holman}, \&
  {Nesvorn{\'y}}}]{2014ApJ...787..132D}
{Deck}, K.~M., {Agol}, E., {Holman}, M.~J., \& {Nesvorn{\'y}}, D. 2014, \apj,
  787, 132, \dodoi{10.1088/0004-637X/787/2/132}

\bibitem[{{Eastman} {et~al.}(2013){Eastman}, {Gaudi}, \&
  {Agol}}]{2013PASP..125...83E}
{Eastman}, J., {Gaudi}, B.~S., \& {Agol}, E. 2013, \pasp, 125, 83,
  \dodoi{10.1086/669497}

\bibitem[{{Foreman-Mackey} {et~al.}(2013){Foreman-Mackey}, {Hogg}, {Lang}, \&
  {Goodman}}]{2013PASP..125..306F}
{Foreman-Mackey}, D., {Hogg}, D.~W., {Lang}, D., \& {Goodman}, J. 2013, \pasp,
  125, 306, \dodoi{10.1086/670067}

\bibitem[{{Fulton} {et~al.}(2018){Fulton}, {Petigura}, {Blunt}, \&
  {Sinukoff}}]{2018PASP..130d4504F}
{Fulton}, B.~J., {Petigura}, E.~A., {Blunt}, S., \& {Sinukoff}, E. 2018, \pasp,
  130, 044504, \dodoi{10.1088/1538-3873/aaaaa8}

\bibitem[{{Hunter}(2007)}]{matplotlib}
{Hunter}, J.~D. 2007, Computing in Science and Engineering, 9, 90,
  \dodoi{10.1109/MCSE.2007.55}

\bibitem[{{Lovis} \& {Fischer}(2010)}]{2010exop.book...27L}
{Lovis}, C., \& {Fischer}, D. 2010, in Exoplanets, ed. S.~{Seager}, 27--53

\bibitem[{McKinney(2010)}]{pandas}
McKinney, W. 2010, in Proceedings of the 9th Python in Science Conference, ed.
  S.~van~der Walt \& J.~Millman, 51 -- 56

\bibitem[{{Murray} \& {Correia}(2010)}]{2010exop.book...15M}
{Murray}, C.~D., \& {Correia}, A.~C.~M. 2010, in Exoplanets, ed. S.~{Seager},
  15--23

\bibitem[{{Murray} \& {Dermott}(1999)}]{1999ssd..book.....M}
{Murray}, C.~D., \& {Dermott}, S.~F. 1999, {Solar system dynamics}

\bibitem[{{Paddock}(1913)}]{1913PASP...25..208P}
{Paddock}, G.~F. 1913, \pasp, 25, 208, \dodoi{10.1086/122237}

\bibitem[{{Tremaine}(2023)}]{2023dyps.book.....T}
{Tremaine}, S. 2023, {Dynamics of Planetary Systems}

\bibitem[{{van der Walt} {et~al.}(2011){van der Walt}, {Colbert}, \&
  {Varoquaux}}]{numpy}
{van der Walt}, S., {Colbert}, S.~C., \& {Varoquaux}, G. 2011, Computing in
  Science and Engineering, 13, 22, \dodoi{10.1109/MCSE.2011.37}

\end{thebibliography}
\bibliographystyle{aasjournal}
\end{document}